# Galilean Invariant Fluid-Solid Interfacial Dynamics in Lattice Boltzmann Simulations

Binghai Wen[1,2,3], Chaoying Zhang[3], Yusong Tu[4], Chunlei Wang[1], Haiping Fang[1,*]

[1] Shanghai Institute of Applied Physics, Chinese Academy of Sciences, Shanghai 201800, China

[2] University of Chinese Academy of Sciences, Beijing 100049, China

[3] College of Computer Science and Information Engineering, Guangxi Normal University, Guilin 541004, China

[4] Institute of Systems Biology, Shanghai University, Shanghai 200444, China

**Galilean invariance is a fundamental property; however, although the lattice Boltzmann equation itself is Galilean invariant, this property is usually not taken into account in the treatment of the fluid-solid interface. Here, we show that consideration of Galilean invariance in fluid-solid interfacial dynamics can greatly enhance the computational accuracy and robustness in a numerical simulation. Surprisingly, simulations are so vastly improved that the force fluctuation is very small and a time average becomes unnecessary.**

*PACS number: 47.11.-j; 47.11.Qr; 05.10.-a*

[*] Corresponding author. Email: fanghaiping@sinap.ac.cn

# I. INTRODUCTION

In the governing equations in numerical simulations, Galilean invariance is usually guaranteed from a fundamental perspective. The original lattice gas automaton (LGA) [1] was not Galilean invariant due to the presence of a nonphysical coefficient in the nonlinear advection term [2]. The lattice Boltzmann equation (LBE) [3-6] eliminated this artifact with the use of a proper equilibrium distribution function in the collision term, and in some investigations a few high-order, even complete Galilean invariant LBE models have been achieved [7-9]. Nowadays, LBE is particularly successful in simulations involving interfacial dynamics [10-13], microflows [14, 15], multiphase flows [16-18], and complex fluid flows [19-21]. However, Galilean invariance in the treatment of the fluid-solid interface has received little attention although it is well known that the boundary has a major influence on the fluid flow. For example, the widely used momentum exchange method does not satisfy Galilean invariance [22-25], and this may be why the method does not have very high computational accuracy.

In the lattice Boltzmann method (LBM), the momentum transfer across a given interface can be computed effectively with the discrete momentum component, and the hydrodynamic force is evaluated easily using the momentum exchange method. Ladd [26] defined originally the suspending particle as a shell with interior fluids and the momentum transfer across the particle boundary is obtained considering the inside and outside fluids separately. This work promoted the lattice Boltzmann method to become a popular tool in simulating fluid-solid interaction problems. Aidun *et al*. [27]

directly represented the impermeable particle without an interior fluid by using a modified half-way bounce-back boundary condition, and the solid-to-fluid density ratio can be regulated freely. Mei *et al.* [28] applied a curved boundary condition to evaluate the hydrodynamic force on the real particulate geometry. Aidun *et al.* [27], Huang *et al.* [29] and Wen *et al.* [30] further considered the additional momenta induced by the type-changing lattices. In these advances, the conventional equation of the momentum exchange methods remains the same, but Galilean invariance is not guaranteed in the treatment of the fluid-solid interface yet [22-25]. Recently, Caiazzo *et al.* [22] and Lorenz *et al.* [24] introduced a correction term to improve the Galilean invariance of the momentum exchange method. Clausen *et al.* [23] proposed a correction to reduce the error of normal stress and investigated the effect on the rheological properties in particle suspensions. Zhou *et al.* [25] coupled the Lees-Edwards boundary condition with a node-based method to studied particle-fluid suspensions. Although the numerical errors caused by non-Galilean effects are significantly diminished, attempts to achieve full Galilean invariance along with high simulating accuracy have not been satisfactory [24, 25].

In this paper, we present a Galilean invariant momentum exchange equation by introducing the relative velocity into the interfacial momentum transfer to compute the hydrodynamic force. The algorithm is simple and independent of boundary geometries. It is demonstrated to ensure Galilean invariance and achieve high accuracy in the dynamic fluid. Remarkably, the consideration of Galilean invariance can greatly enhance the computational accuracy and robustness of fluid-solid interfacial dynamics, so that the widely used time averaged computation of velocity and force becomes unnecessary.

## II. LATTICE BOLTZMANN METHOD

With its roots in kinetic theory and the cellular automaton concept, the lattice Boltzmann equation can obtain the incompressible Navier-Stokes equations in the nearly incompressible limit [2, 31, 32]. Discretized fully in space, time and velocity, the lattice Boltzmann equation can be concisely written as

$$f_i(\bm{x}+\bm{e}_i,t+1) - f_i(\bm{x},t) = \bm{\Omega}(f_i), \tag{1}$$

where $f_i(\bm{x},t)$ is the particle distribution function at lattice site $\bm{x}$ and time $t$, moving along the direction defined by the discrete speeds $\bm{e}_i$ with $i = 0, ..., N$, and $\bm{\Omega}(f_i)$ is the collision operator. With the different collision operators, several variations of the LBE can be read as the single-relaxation-time (SRT) mode [3-6], the multiply-relaxation-time (MRT) model [33, 34], the two-relaxation-time (TRT) model [35], the entropic lattice Boltzmann equation (ELBE) [36, 37], etc. The mass density and the momentum density are defined by $\rho = \sum f_i$ and $\rho\bm{u} = \sum \bm{e}_i f_i$, respectively. One can consider $f_i$ to be a mass component of a lattice node, and $\bm{e}_i f_i$ to be the corresponding momentum component. The evolution of the LBE can be decomposed into two elementary steps, collision and advection:

$$\text{collision:} \quad \widetilde{f}_i(\bm{x},t) = f_i(\bm{x},t) + \bm{\Omega}(f_i), \tag{2}$$

$$\text{advection:} \quad f_i(\bm{x}+\bm{e}_i,t+1) = \widetilde{f}_i(\bm{x},t), \tag{3}$$

where $f_i$ and $\tilde{f}_i$ denote pre-collision and post-collision states of the particle distribution functions, respectively. The dominant part of the computations, namely the collision step, is completely local, hence the discrete equation is natural to parallelize.

## III. GALILEAN INVARIANT MOMENTUM EXCHANGE METHOD

### A. Conventional momentum exchange equation

The interfacial momentum transfer in the conventional momentum exchange methods (CME) [26-30] can be generalized by a common schematic diagram as shown in Fig. 1(a). A moving boundary is located between a fluid node $x_f$ and a boundary node $x_b$ (it is an interior fluid node in the method of Ladd [26]). The boundary has a vector velocity $v$ at the point of intersection $S$. In the collision step, the distribution function $\tilde{f}_{\bar{i}}(x_b, t)$ can be calculated by the interior fluid evolution [26], the half-way bounce-back boundary condition [27] or the curved boundary conditions [28-30], in which the forcing terms [38-40] based on the boundary velocity need to be included.

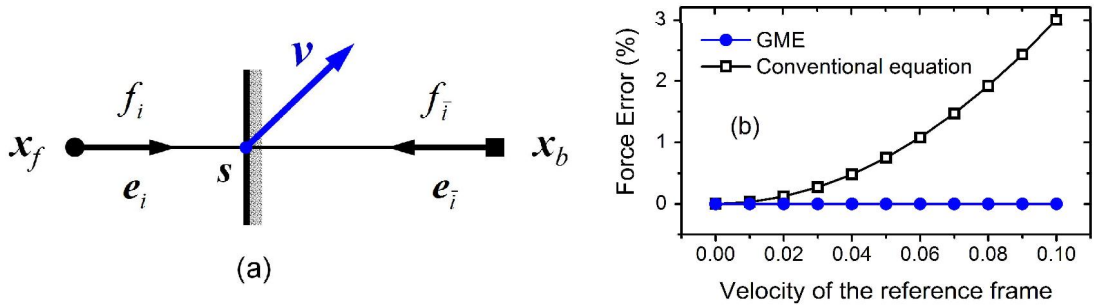

Fig. 1 (color online). (a) A common schematic diagram to illustrate a moving boundary crossing a fluid-solid link at the point of intersection S. $x_f$ and $x_b$ denote the

*adjacent fluid and boundary nodes. The boundary has a velocity **v** at the point **S**. (b) Relative errors in the one-sided pressure on a vertical thin plate in the relatively stationary fluid without boundaries. This equilibrium system is connected to various velocities of the reference frame. The conventional equation, i.e., Eq. (4), obviously violates Galilean invariance. Since it properly considers the boundary velocity, the present method (GME), i.e., Eq. (5), is fully Galilean invariant and thus has a very high computational accuracy.*

When distribution functions propagate, the mass component $\tilde{f}_i(\boldsymbol{x}_f,t)$ streams into the boundary and contributes a momentum increment, while $\tilde{f}_{\bar{i}}(\boldsymbol{x}_b,t)$ streams out of the boundary and contributes a momentum decrement. In the literature [26-28, 30], these momentum components are calculated by directly using the discrete velocities $\boldsymbol{e}_i$ and $\boldsymbol{e}_{\bar{i}}$, namely $\boldsymbol{e}_i\tilde{f}_i(\boldsymbol{x}_f,t)$ and $\boldsymbol{e}_{\bar{i}}\tilde{f}_{\bar{i}}(\boldsymbol{x}_b,t)$. The conventional equation for the momentum exchange methods to evaluate the force on a fluid-solid link can be written as [26-28, 30]

$$\boldsymbol{F}(\boldsymbol{x}_s) = \boldsymbol{e}_i\tilde{f}_i(\boldsymbol{x}_f,t) - \boldsymbol{e}_{\bar{i}}\tilde{f}_{\bar{i}}(\boldsymbol{x}_b,t). \tag{4}$$

Although a few modifications have been proposed to improve the accuracy and Galilean invariance [22-24], the concept of the conventional equation remains the same all the time.

## B. Galilean invariant momentum exchange method

It is clear that the momentum component $e_i \widetilde{f}_i$ uses the lattice as the frame of reference, and Eq. (4) is unrelated to the boundary velocity $v$. Considering that relative velocity is used in the momentum theorem, Eq. (4) makes an implicit assumption that the boundary would be motionless during the momentum transfer, regardless of the speeds of the reference frame and the actual boundary. This assumption obviously violates Galilean invariance and causes a divergent difference, which can be expressed numerically in Fig. 1(b) and analytically in an equilibrium system in Part C.

However, the momentum transfer is correlated to the relative velocity and is independent of the frame of reference. When a distribution function propagates across the boundary, the relative velocity at the intersection point should be used in the momentum computation. Crossing the point of intersection $S$, the mass component $\widetilde{f}_i(x_f, t)$ has the velocity $(e_i - v)$ relative to the boundary and it contributes a momentum increment $(e_i - v)\widetilde{f}_i(x_f, t)$ to the boundary. Simultaneously, the mass component $\widetilde{f}_{\bar{i}}(x_b, t)$ has the relative velocity $(e_{\bar{i}} - v)$ and decreases a momentum $(e_{\bar{i}} - v)\widetilde{f}_{\bar{i}}(x_b, t)$ from the boundary. According to the theorem of momentum, the Galilean invariant momentum exchange method (GME) can be defined by

$$F(x_s) = (e_i - v)\widetilde{f}_i(x_f, t) - (e_{\bar{i}} - v)\widetilde{f}_{\bar{i}}(x_b, t) \tag{5}.$$

The total hydrodynamic force $F$ and torque $T$ acting on the solid particle are evaluated in the same way as the convention methods [28, 30]

$$F = \sum F(x_s) \tag{6}$$

and

$$T = \sum (x_s - R) \times F(x_s), \tag{7}$$

where $R$ is the mass center of the solid particle, and the summation runs over all the fluid-solid links.

The momentum components used in the force evaluation are always on the fluid-solid links and GME turns into CME when the boundary is motionless, GME therefore is consistent to the previous theoretical analysis [41]. It should be noted that the consideration of the boundary velocity in Eq. (5) is different from the forcing terms [38-40] in the moving boundary conditions [40-44]. A forcing term, which contains a boundary velocity, represents the effect that the moving boundary exerts on the bounced-back distribution functions, whereas the present method use the boundary velocity to compute the momentum transfer in terms of the momentum theorem. GME evaluates the hydrodynamic force in the fluid-solid interaction and works on the motion state of moving boundaries, but has not any direct influence on distribution functions.

### C. Comparisons in equilibrium state

We employ a simple analysis to compare straightforward the present equation and the conventional one. Suppose both of the fluid and the boundary in Fig. 1(a) have an arbitrary uniform velocity $v$, then the system is physically related to a frame of reference with the uniform velocity $-v$ and is equivalent to a quiescent system. As the system remains in the equilibrium state, the distribution functions are always equal to the equilibrium functions. Let us use the equilibrium distribution function [2]

$$f_i^{(eq)} = \rho\omega_i[1+3(\boldsymbol{e}_i\cdot\boldsymbol{u})+\frac{9}{2}(\boldsymbol{e}_i\cdot\boldsymbol{u})^2-\frac{3}{2}u^2], \tag{8}$$

where $\omega_i$ is the weighting coefficient and $\boldsymbol{u}$ is the fluid velocity, the hydrodynamic force on a fluid-solid link can be obtained according to Eqs. (4) and (8)

$$\begin{aligned}\boldsymbol{F}_i &= \boldsymbol{e}_i f_i^{(eq)}(\boldsymbol{x}_f,t)-\boldsymbol{e}_{\bar{i}} f_{\bar{i}}^{(eq)}(\boldsymbol{x}_b,t) \\ &= \boldsymbol{e}_i\rho\omega_i[1+3(\boldsymbol{e}_i\cdot\boldsymbol{v})+\frac{9}{2}(\boldsymbol{e}_i\cdot\boldsymbol{v})^2-\frac{3}{2}v^2]-\boldsymbol{e}_{\bar{i}}\rho\omega_{\bar{i}}[1+3(\boldsymbol{e}_{\bar{i}}\cdot\boldsymbol{v})+\frac{9}{2}(\boldsymbol{e}_{\bar{i}}\cdot\boldsymbol{v})^2-\frac{3}{2}v^2] \\ &= 2\rho\omega_i\boldsymbol{e}_i+3\rho\omega_i[3(\boldsymbol{e}_i\cdot\boldsymbol{v})^2-v^2]\boldsymbol{e}_i.\end{aligned} \tag{9}$$

Because of the term $3\rho\omega_i[3(\boldsymbol{e}_i\cdot\boldsymbol{v})^2-v^2]\boldsymbol{e}_i$, the resulting force changes abnormally with the speed of the reference frame. Hence, the conventional equation obviously presents an inherent flaw of Galilean invariance, and the difference is in proportion to the modulus square of the reference velocity.

As the discrete velocity $\boldsymbol{e}_i$ is constant in the LBE, Galilean invariance cannot be satisfied on a single fluid-solid link, just like a single $\boldsymbol{e}_i$ cannot express the fluid velocity of a lattice node. However, the discrete velocity set is symmetrical, so that the Galilean invariant force evaluation can be achieved locally on the lattice. Using the D2Q9 model with the discrete velocity set $\boldsymbol{e}=\{(0,0),(1,0),(0,1),(-1,0),(0,-1),(1,1),(-1,1),(-1,-1),(1,-1)\}$, without loss of generality, we assume that the boundary intersects with $\boldsymbol{e}_2$, $\boldsymbol{e}_5$, and $\boldsymbol{e}_6$. In the equilibrium system above, the local hydrodynamic forces on the three fluid-solid links are calculated analytically according to Eqs. (5) and (8)

$$\begin{aligned}
\boldsymbol{F} &= \boldsymbol{F}_2 + \boldsymbol{F}_5 + \boldsymbol{F}_6 \\
&= \sum_{i=2,5,6}[(\boldsymbol{e}_i - \boldsymbol{v})f_i^{(eq)}(\boldsymbol{x}_f,t) - (\boldsymbol{e}_{\bar{i}} - \boldsymbol{v})f_{\bar{i}}^{(eq)}(\boldsymbol{x}_{bi},t)] \\
&= \sum_{i=2,5,6}\{2\rho\omega_i\boldsymbol{e}_i + 3\rho\omega_i[3(\boldsymbol{e}_i \cdot \boldsymbol{v})^2 - \boldsymbol{v}^2]\boldsymbol{e}_i - 6\rho\omega_i(\boldsymbol{e}_i \cdot \boldsymbol{v})\boldsymbol{v}\} \\
&= \sum_{i=2,5,6}2\rho\omega_i\boldsymbol{e}_i.
\end{aligned} \quad (10).$$

With the simple vector calculation, the term related to the reference velocity *v*, as shown in Eq. (9), is eliminated due to the symmetry of the velocity set. The local hydrodynamic force remains constant regardless of the speed of the reference velocity; thus, GME is proven to be completely Galilean invariant in the equilibrium system. The simplest case that shows the difference between GME and the conventional equation is a computation of the one-sided pressure on a vertical thin plate, which is placed in the relatively static fluid without boundaries. The equilibrium system is connected to a horizontal reference speed and the benchmark is computed in the quiescent system. Fig. 1(b) compares the percentage of the computational errors in the hydrodynamic forces computed by GME and the conventional equation. The case is independent of the relaxation time and the plate length. It is clear that the conventional equation violates Galilean invariance whereas GME fully satisfies in the equilibrium system.

## IV. SIMULATION RESULTS AND DISCUSSION

Particle suspension is a very effective way to investigate the accuracy and Galilean invariant of force evaluation. Since the particles in our test cases are unconfined freely moving cylinder and sphere under the combined action of gravity and hydrodynamic force, the errors of the forces will be accumulated and then be displayed apparently. In this section, we deeply investigate the accuracy, robustness and Galilean invariance of GME by a series of direct numerical simulations, in which part (A), (B) and (C) are two-dimensional cylinder sedimentations and part (D) is a three-dimensional rigid sphere migrating laterally in a Poiseuille flow. The simulations apply the second-order interpolation boundary condition [40] on the SRT model with the single relaxation time $\tau = 0.6$. The highly consistent results are obtained by using the multireflection boundary condition [41] on the MRT model with the diagonal relaxation matrix $\hat{\boldsymbol{S}} = \text{diag}(0, 1.64, 1.54, 0, 1.9, 0, 1.9, 1/\tau, 1/\tau)$ [33, 45].

### A. Galilean invariance in dynamic system

We demonstrate Galilean invariance and the computational accuracy of the present scheme in a dynamic system by examining cylinder sedimentations [46]. As shown in Fig. 2, a cylinder is initially released away from the centerline of a vertical channel with static fluid. Since the mass density of the cylinder is somewhat bigger than the fluid's, it rotates and translates under the gravitational and hydrodynamic forces. Finally, it reaches a steady state descending along the centerline at a constant velocity. The channel width is 0.4 cm and the cylinder diameter is 0.1 cm. The fluid density and kinematic viscosity are 1 g/cm$^3$ and 0.01 cm$^2$/s. The cylinder is released at 0.076 cm away from the left wall, and then it settles under the gravity acceleration |$\boldsymbol{G}$|=980 cm$^2$/s. The width of the channel is 120 lattice units and the length is 10 times the width.

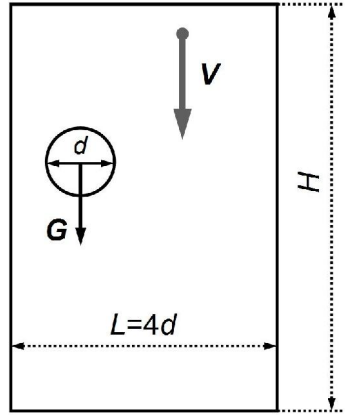

*Fig 2. A schematic diagram of cylinder sedimentation in a vertical channel, **G** is the gravity and **V** is the velocity of the reference frame.*

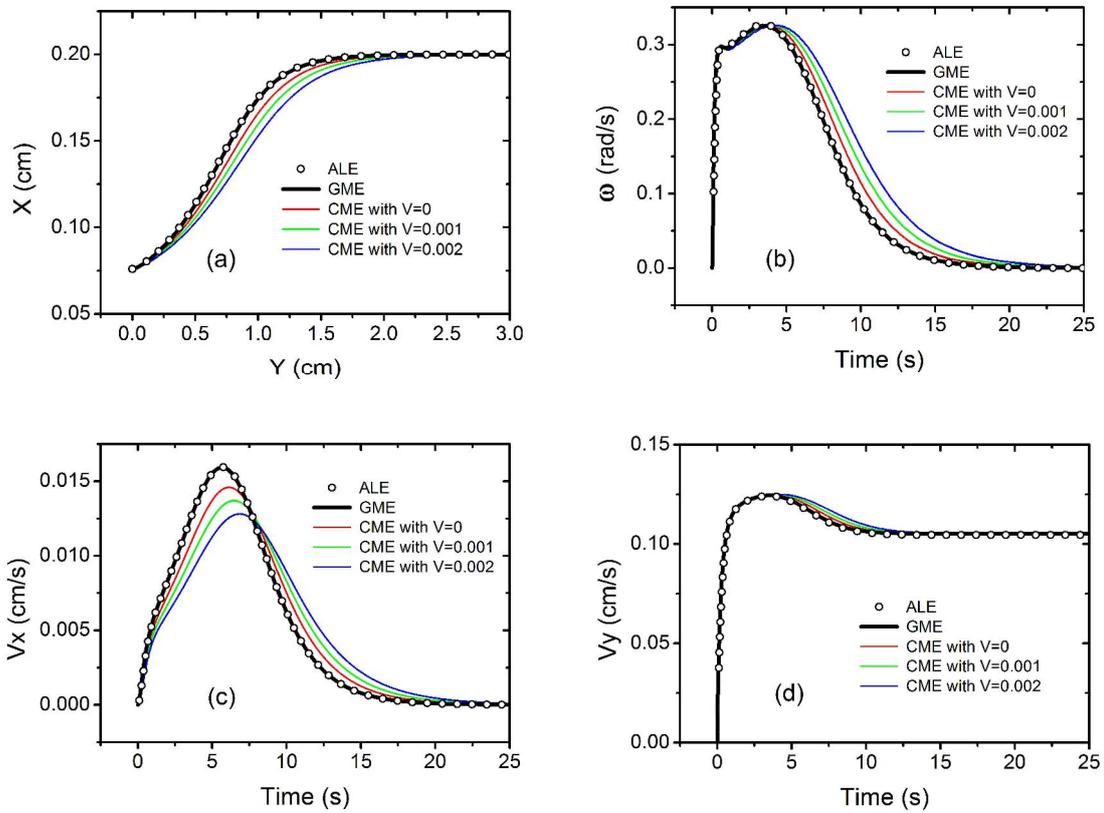

*Fig. 3 (color online). Time-dependent (a) trajectories, (b) angular velocities, (c) horizontal velocities and (d) vertical velocities relative to the channel. The density of the cylinder is 1.003 g/cm$^3$ and the terminal Reynolds numbers is 1.03. The dynamic*

simulation system is connected to three velocities of the reference frame, i.e., *V=0, 0.001, and 0.002.*

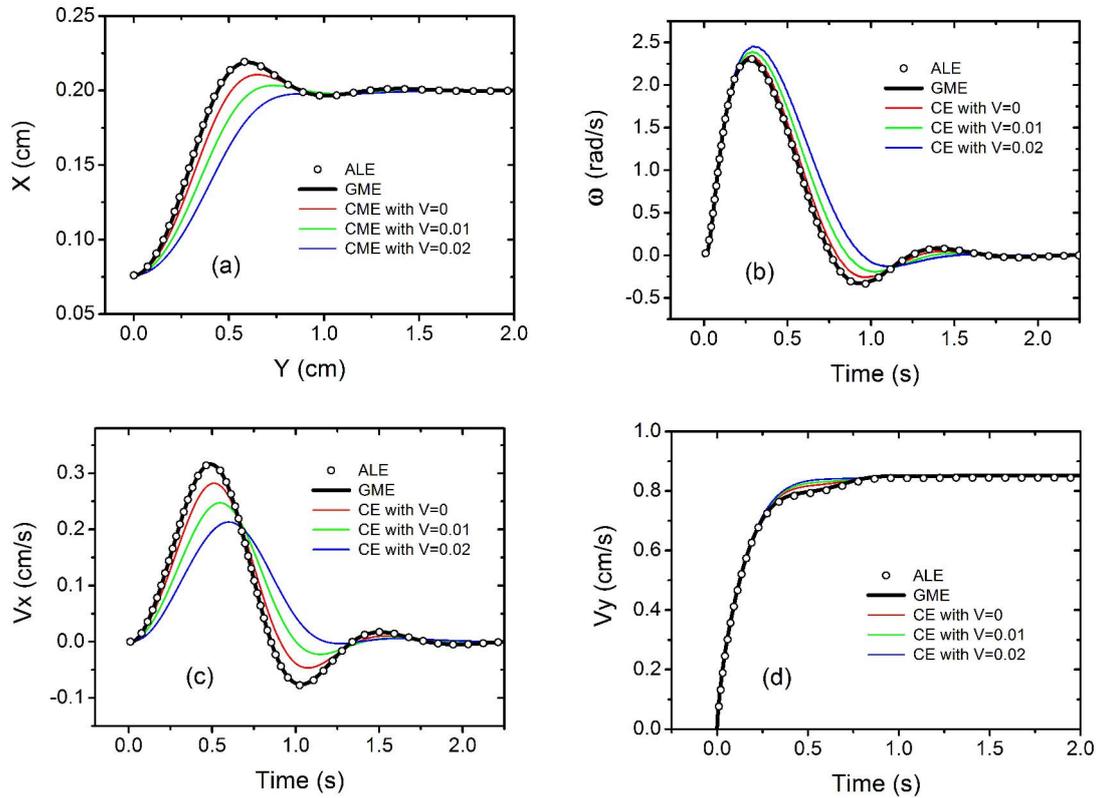

Fig. 4 (color online). Time-dependent (a) trajectories, (b) angular velocities, (c) horizontal velocities and (d) vertical velocities relative to the channel. The density of the cylinder is 1.03 g/cm$^3$ and the terminal Reynolds numbers is 8.33. The dynamic simulation system is connected to three velocities of the reference frame, i.e., *V=0, 0.01, and 0.02.*

The densities of the cylinder in two simulations are 1.003 and 1.03 g/cm$^3$, respectively. The terminal Reynolds numbers of the particles are 1.03 and 8.33 correspondingly, which is defined by $Re = du_p/v$, where $u_p$ is the final velocity of the particle and $\nu$ is the kinematic viscosity. We place the simulation system in several uniform frames of reference. Explicitly, we initially assign additional uniform velocities to the fluid, the particle and the channel, $V = 0, 0.001, 0.002$ for the former and $V = 0, 0.01, 0.02$ for the latter, respectively. The time-dependent trajectories, angular velocities, horizontal velocities and vertical velocities relative to the channel are presented in Fig. 3 and 4 together with the comparison with the results by the conventional equation and the arbitrary Lagrangian–Eulerian technique (ALE) [46]. The results of the conventional equation show sizeable differences from the benchmarks, even if the reference frame is stationary. And the deviations grow larger and larger as the reference velocities increase. These indicate that the conventional equation is not suitable for moving boundaries. However, regardless of the speed of the reference velocities, the GME results are always in excellent agreement with the benchmarks. These numerical simulations support that GME meet a full Galilean invariance, and therefore we draw only one line to represent the GME results with the various reference speeds.

### B. Accuracy of hydrodynamic force

Now, without any reference velocity, we consider the sedimentation of the cylinder with 1.03 g/cm$^3$ to demonstrate the vast improvement in the computational accuracy by using the present method. Figs. 5(a) and (b) draw the compare with the simulating results from the previous momentum exchange methods, ALD [27] and LME [30], together with the benchmarks from ALE. The hydrodynamic forces computed by GME extremely agree with the benchmarks, while the results by ALD and LME have large fluctuations. Please note that all of the data from GME are raw, whereas the data from ALD and LME have been smoothed using the adjacent-averaging method — per 30 points for the horizontal forces and per 100 points for the vertical forces. As the improvement in the force evaluation is so great, the force fluctuation of GME is very small and the time average becomes unnecessary.

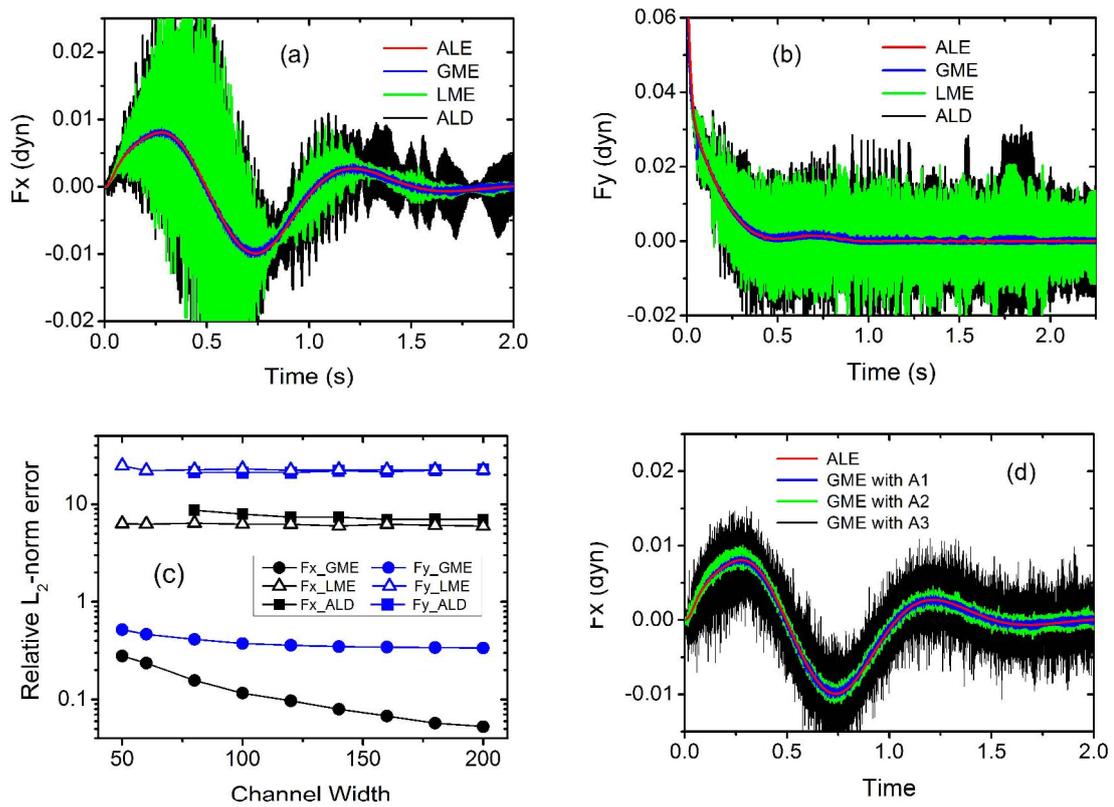

*Fig. 5 (color online). (a) Time-dependent horizontal forces and (b) time-dependent vertical forces evaluated by GME, LME, and ALD, compared with the ALE benchmark. The density of the cylinder is 1.03 g/cm$^3$. The GME data is raw, whereas the ALD and LME data have been smoothed by the adjacent-averaging method. (c) The relative L$_2$-norm error of the horizontal forces (Fx, black) and vertical forces (Fy, blue) under increasing lattice scales. (d) GME simulations coupled with the different*

*algorithms to fill newborn fluid nodes, second-order extrapolation (A1), linear extrapolation (A2) and neighbor-node average (A3).*

We carefully compare the effect of the lattice scale for different schemes of force evaluations by performing a set of simulations in which the various lattice sizes are used to simulate the same cylinder sedimentation with the particle density 1.03 g/cm$^3$. The lattice width of the channel increases gradually from 50 to 200 lattice units, while the length remains 10 times the width. The degree of force deviation is indicated by the relative L$_2$-norm error, which is defined by

$$E = \frac{\{\int [f(t) - F(t)]^2 dt\}^{1/2}}{\{\int [F(t)]^2 dt\}^{1/2}}, \qquad (11)$$

where *f(t)* is a LBM result and *F(t)* is an ALE result. Fig. 5(c) illustrates that the relative errors of the GME results rapidly decrease with the increase of the lattice scale. However, the relative errors of the ALD and LME results always remain very high and are more than one order larger than those from GME.

We emphasize that the small fluctuations of the GME data are mostly unrelated to Eq. (5); they are mainly caused by the inaccurate distribution functions of the newborn fluid nodes and have the potential to be reduced further. Using the fluid nodes on the around fluid-solid links, three straightforward algorithms are employed to fill the newborn fluid nodes [30, 47], namely second-order extrapolation [40] (A1), linear extrapolation (A2) and neighbor-node average (A3). If the participant fluid-solid links are more than one, the newborn is assigned as their average. It is evident in Fig. 5(d) that a good algorithm can remarkably reduce the fluctuations. The A1 algorithm is also used in all other simulations in the present work. Please refer to the papers [24, 40, 47] to know more useful algorithms about the issue.

### C. Fluctuations of velocity and angle velocity

We further analysis the accuracy of the velocities and the angle velocities by GME, ALD and LME. As shown in Figs. 6(a), (b) and (c), all velocities from GME are very smooth and in excellent agreement with the ALE benchmarks, whereas the results from ALD and LME clearly fluctuate with some deviations. The density of the cylinder is 1.03 g/cm$^3$ in the simulations.

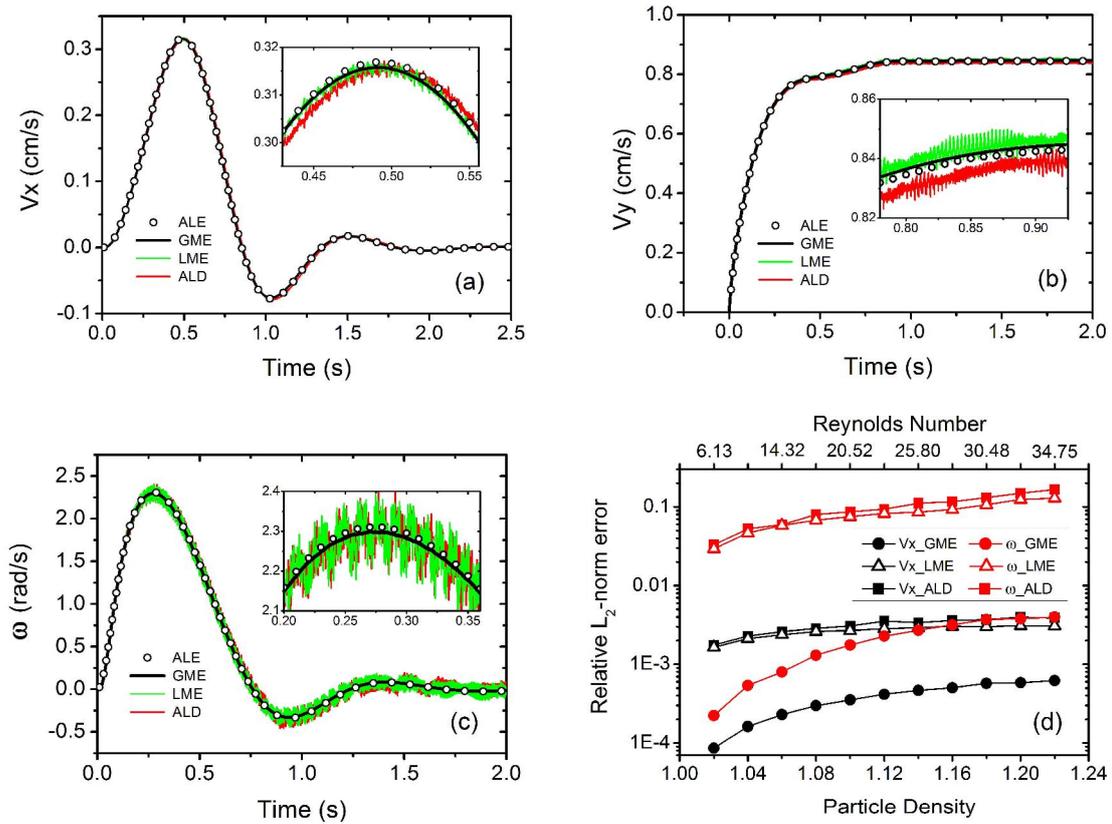

Fig. 6 (color online). (a) Time-dependent horizontal velocities, (b) time-dependent vertical velocities, and (c) time-dependent angular velocities evaluated by GME, LME,

*and ALD, compared with the ALE benchmark. (d) The relative L$_2$-norm errors of the horizontal velocity (Vx, black) and the angular velocity (ω, red) with various particle densities.*

To investigate the influence of the Reynolds number, a set of simulations are performed with the different particle densities which increase from 1.02 to 1.22 g/cm$^3$. The moderate Reynolds number is defined by $Re = du/\nu$, where $d$ is the cylinder diameter, $u$ is the final velocity of the particle and $\nu$ is the kinematic viscosity. In these simulations, the Reynolds number grows gradually from 6.13 to 34.75. The degree of fluctuation is also indicated by the relative L$_2$-norm error, where *f(t)* is the simulation result and *F(t)* is the smoothed result by the adjacent-averaging method per 20 points. It is clearly shown in Fig. 6(d) that the GME results are more accurate and far steadier than the ALD and LME results and that the time average of the velocities is totally unnecessary.

### D. Three dimensional numerical simulation

The Galilean invariant momentum exchange method can be easily extended to three-dimensional systems. We perform the simulations of a neutrally buoyant rigid sphere migrating laterally in a tube Poiseuille flow, which is schematically illustrated in Fig. 7. This phenomena is called the Segré-Silberberg effect and was discovered in 1962 [48]. The tube radius is 0.2 cm and the sphere radius is 0.061 cm. The fluid density is 1.05 g/cm$^3$, the dynamic viscosity is 1.2 poise and the flow rate is 0.0711 cm³/s. In the present simulations, the sphere radius is 5.9 lattice units and the length of the tube is 150 lattice units. The pressure drops from the inlet to the outlet is 1.825E-5 and pressure boundary condition [49] is applied at both the inlet and outlet of the tube.

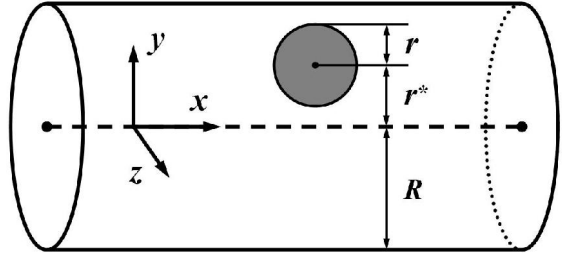

Fig 7. A schematic diagram of a neutrally buoyant sphere migrating in a tube Poiseuille flow.

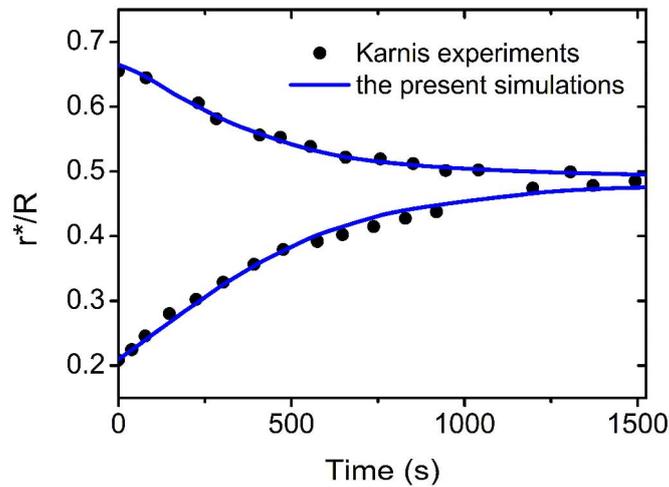

Fig. 8 (color online). Three-dimensional simulations of the Segré-Silberberg effect by the lattice Boltzmann equation with GME.

Fig. 8 presents two trajectories of the spheres released at the dimensionless radial positions of $r^*/R=0.21$ and 0.66, where $r^*$ is the radial distance from the tube centerline. Different from the 2D results [30], the equilibrium positions of the spheres are far from the centerline. The numerical results by the lattice Boltzmann simulations with GME are highly consistent with the experiments by Karnis *et al*. [50]. This verifies that GME is competent to three-dimensional dynamic simulations.

## V. CONCLUSION

In this work we propose a Galilean invariant momentum exchange equation to compute the hydrodynamic force by introducing the relative velocity into the interfacial momentum transfer. Numerical cases support strongly that the scheme meet full Galilean invariance. We further find that Galilean invariance is not only a basic rule, but also plays a key role in improving the numerical accuracy in lattice Boltzmann simulations. Direct numerical simulations of the cylinder sedimentations and the three-dimensional Segré-Silberberg effect confirm that GME is able to exactly depict the behaviors of suspension particle and holds an excellent stability. The present algorithm only uses local data and is independent of boundary geometries; thus, it is efficient and easily implemented in both two and three dimensions. GME can be combined with many curved boundary conditions [40-44] and be adopted in different lattice Boltzmann models, such as SRT, MRT, TRT, and ELBE. We expect the present method will promote the applications of LBM in various dynamic and complex systems, for example moving vehicles, artery motions, colloidal suspensions, etc.

We thank Prof. Yuehong Qian and Prof. Zhaoli Guo for useful discussions. This work was supported by NSFC (10825520, 11162002 and 11105088), NBRPC (2012CB932400), IPSMEC (11YZ20), and Shanghai Supercomputer Center of China.